# Design and production of the DESI fibre cables

Jürgen Schmoll[a,*], Robert Besuner[c], David Bramall[a], David Brooks[d], Jerry Edelstein[b], Patrick Jelinsky[c], Michael Levi[b], Graham Murray[a], Claire Poppett[b], Ray Sharples[a], Luke Tyas[a], David Schlegel[b] for the DESI collaboration

[a] Centre for Advanced Instrumentation, Netpark Research Institute, Joseph Swan Way, Netpark, Sedgefield TS21 3FB, United Kingdom;
[b] Lawrence Livermore National Laboratory, 1 Cyclotron Road, Berkeley CA94720, USA;
[c] Department of Physics, University of California, Berkeley, CA 94720, USA;
[d] Department of Physics Astronomy, University College London, Gower Street, London WC1E 6BT, UK

## 1. ABSTRACT

The Dark Energy Spectroscopic Instrument (DESI) is under construction to measure the expansion history of the Universe using the Baryonic Acoustic Oscillation technique. The spectra of 35 million galaxies and quasars over 14000 sq deg will be measured during the life of the experiment. A new prime focus corrector for the KPNO Mayall telescope will deliver light to 5000 fibre optic positioners. The fibres in turn feed 10 broad-band spectrographs. We will describe the design and production progress on the fibre cables, strain relief system and preparation of the slit end. In contrast to former projects, the larger scale of production required for DESI requires teaming up with industry to find a solution to reduce the time scale of production as well as to minimise the stress on the optical fibres.

## 2. INTRODUCTION

The Dark Energy Survey Instrument DESI will allow a redshift survey of over 35 million galaxies and quasars, utilising the primary focus of the 4m Mayall telescope of the Kitt Peak National Observatory. Ten petals of 500 fibre positioners each will allow sampling up to 5000 objects at a time, with ten optical fibre cables connecting a bank of ten spectrographs. Each spectrograph consists of three channels split by two dichroic mirrors, covering an overall range between 360 and 980 nanometers. At the telescope primary focus, the round focal plane is split into 10 wedge shaped petals and the 500 probes of each petal are transferred into one spectrograph slit on the other end of the fibre cable, with the spectrographs being located in the former coude focus area of the telescope. Here we describe the design and manufacture of the science cables that are used to guide the light of the 5000 objects down into the spectrographs. The cables are being produced by the Centre for Advanced Instrumentation (CfAI) at Durham University. For an overview of the DESI project see also Martini et al., [1].

## 3. FIBRE CABLE ARCHITECTURE AND PRODUCTION

The cable feed needs to be flexible enough to pass through the energy chains used around the RA and Dec bearings of the Mayall telescope while protecting the fibres securely against damage and stress. The optical fibres from the primary focus meet the main fibre link in a strain relief box at the top of the telescope, where they are connected by fusion splicing (Poppett et al, [2] and [3]). As it can be seen in Figure 1, each spectrograph is connected with an about 44m long main cable that is spliced to the primary focus arrangement, passes around the Mayell telescope's two main bearings and finishes at one of ten spectrograhs after passing through a stationary strain relief box. From this box, a 4.1m long section of the cable guides the light into the spectrograph slits. The long part of the cable which is subject to constant motion while the telescope is pointing and tracking has a length of 38.1m. To compensate for the internal length dispersion that the fibres experience in a bent cable, it is sub-divided into eleven conduits around a tensile element as seen in Figure 2. The tensile element is a 11mm Parafil cable that, made out of rubber coated Kevlar, can bear loads up to 6000kg. 11 M2FX (former "Miniflex") conduits (OD 4mm, ID 2.6mm) are grouped around the Parafil cable and wound along in a spiral manner. This "stranding" assures that the positions of each conduit change from being sometimes on the inner and sometimes on the outer side of a bend, averaging the length dispersion out that appears when a thick cable with multiple conduits in it is bent in a curve. This arrangement reduces fibre stress, and as a further precaution the fibre cable is laid out in a "figure eight" manner when stored in the transport box. The whole cable architecture is optimised to reduce stress, following the method developed by Murray et al [6].

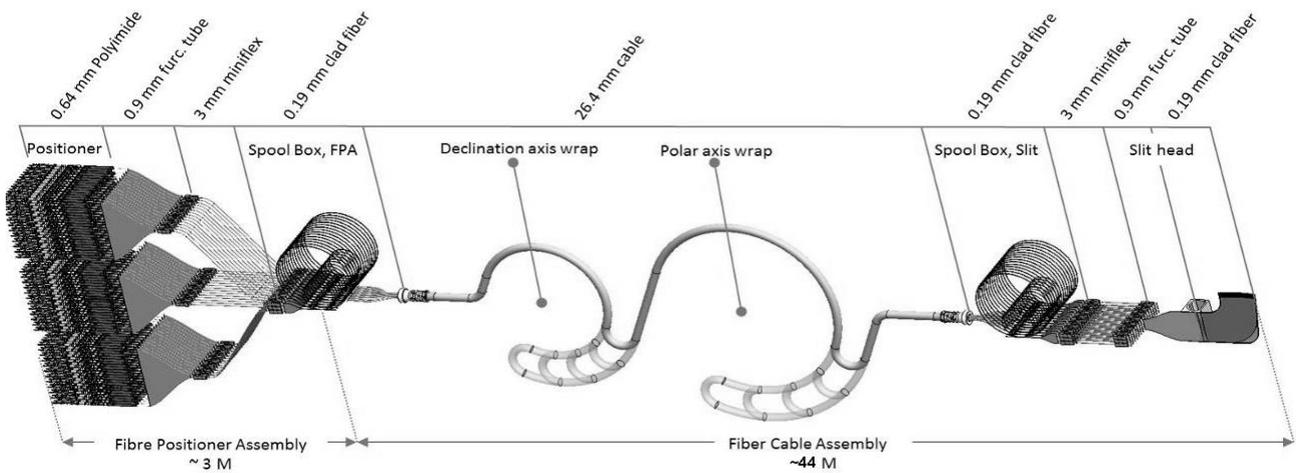

**Figure 1: Schematic layout of the DESI fibre conduit for one science cable feeding 500 fibres to one spectrograph slit.**

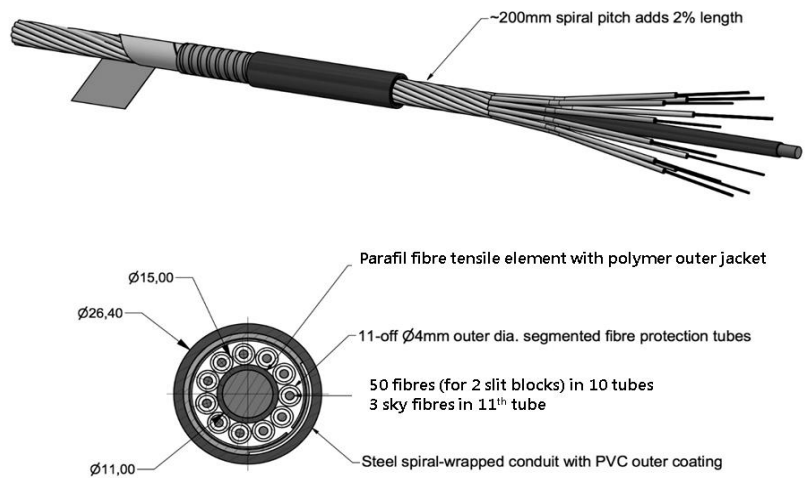

**Figure 2: The architecture of the 38.1m long main cable between the prime focus assembly splice and the spectrograph side strain relief box.**

## 3.1 The cable manufacture: Challenges and solutions

Previous fibre cables were made at the CfAI using conduits with a pull string to pull fibre bundles through that were laid out flat and collated to a bundle. However, the sheer size of the DESI project made this method impractical, as the total fibre length is in excess of 200 kilometers which excluded the standard way of manufacture in a timely manner. In addition, early prototyping of novel pull methods indicated severe problems with focal ratio degradation (FRD) once the fifty fibres were inside the conduit. While one source of the FRD was found in a bad batch of the optical fibre itself (which gave rise to extend quality control to 50m long samples), the other problem could be linked to the way the fifty fibres were drawn into the conduit. As the individual fibres were collated on a grooved spool, various strands kept clinging together and entering the conduit in a rather erratic way. This method caused some fibre fractures, and the FRD of the fibres increased to magnitudes that could not always be measured using the standard collimated beam method (see below) as the expected ring at the fibre output field was hardly recognisable.

This was improved markedly when another insertion method was chosen: Each single fibre was fed from a small individual reel and the fibres were guided through a hole array and a furcation tube mandrel before entering the conduit as shown in Figure 3. Doing it this deterministic way was time consuming (about six hours to pull 50 fibres into one full length test conduit), but the FRD was not significantly higher than for the free strands measured before pulling.

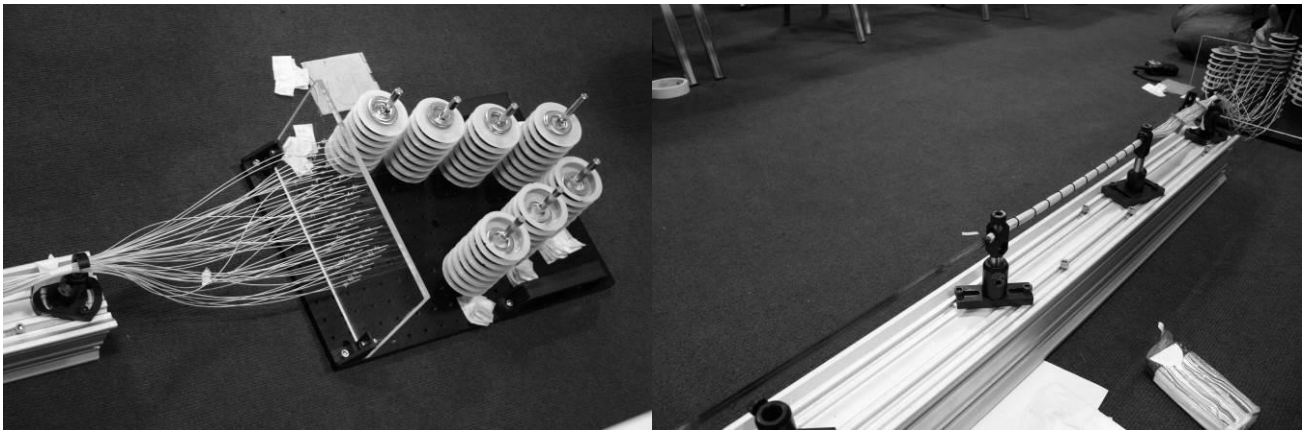

**Figure 3: Experimental 50m fibre cable draw with individual reels and deterministic feed using a hole array and furcation tube mandrel.**

## 3.2 Industrialisation of the conduit manufacture

After the method has been found to feed 50 optical fibres into a Miniflex conduit without increasing the FRD, a streamlined process was necessary to timely deliver the 100 conduits (plus 10 filled with sky fibres) necessary for the project. Collaborating with the conduit manufacturer PPC Miniflex Ltd[1], this method was developed at the premises of the PPC factory.

Replacing the 50m single reels of the experiment with 1200m fibre spools and respooling the fibre batches accordingly, a maximum length of conduit has been produced to be cut into shorter lengths at a later stage.

---

[1] PPC Broadband Fiber Ltd, Unit1, Parham Airfield, Woodbridge IP13 9EZ, United Kingdom

While the fibres were fed into the conduit deterministically using a perforated mask again, the Miniflex conduit get cast around the fibre bundle during the pulling process. This technique allows for pulling an endless quantity at a time without the need to pull the fibre bundle through a laid out conduit. The various steps can bee seen in Figure 4.

The final step for the cable itself is the stranding. The stranding method bases on in-house development for the FMOS and PFS projects (Murray et al, [5], [6]). Using a planetary stranding machine, 11 conduits (10 for the science cables, 1 for three sky fibres) are wound together before a thin plastic tape finishes it off. After cutting the cables to their final length, they are pulled into 38.1m long Adaptaflex conduits so the cable architecture seen in Figure 2 is achieved. At this stage the cable is ready for the strain relief box fitting.

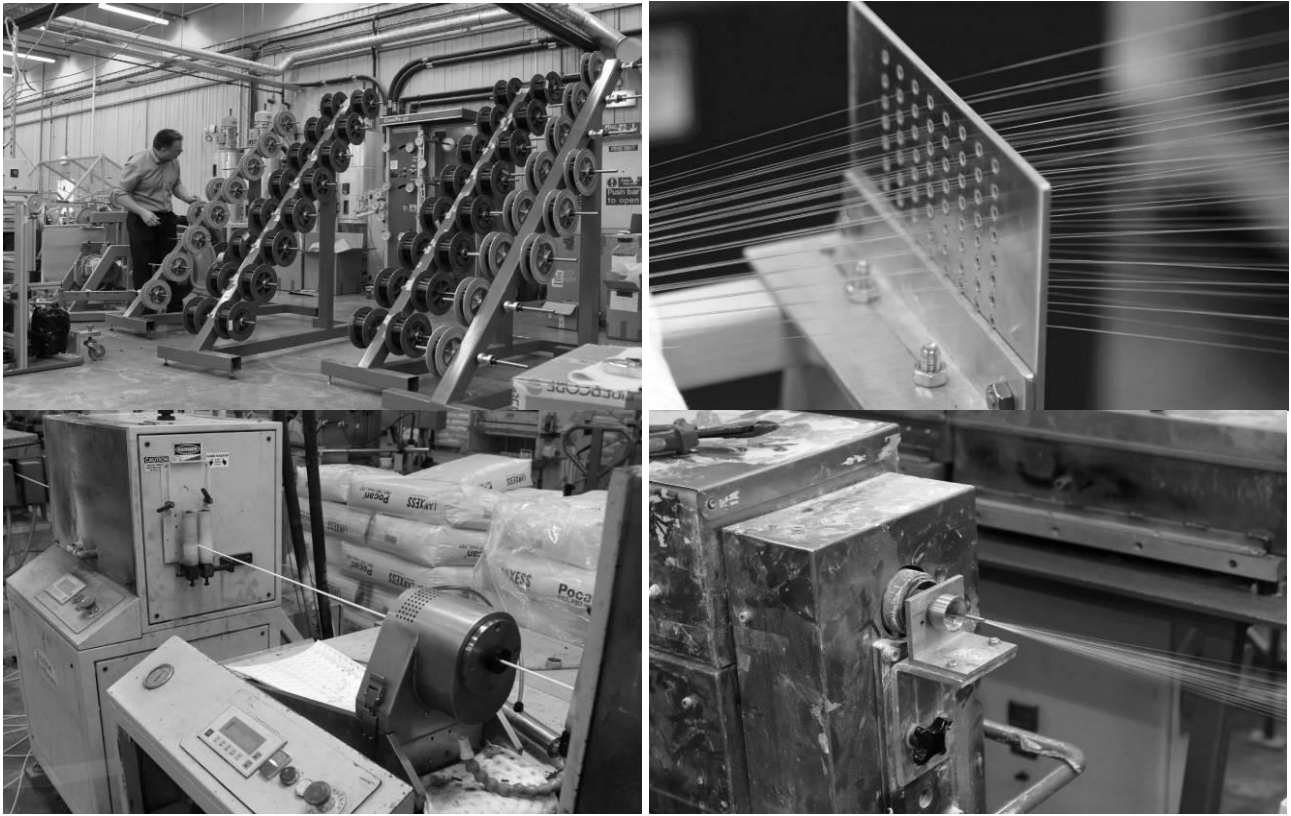

**Figure 4: Manufacture of the Miniflex conduits with insertion of the 50+ optical fibres. Top left: 53 reels of re-spooled fibres; top right: Hole array to guide fibres into mandrel; bottom right: Fibres entering the mandrel; bottom left: Fresh Miniflex tubing with fibres already inserted.**

## 3.3 Strain relief measures

Despite the efforts of stranding the cables and storing them in "figure eight" windings, there are still small differences in length that vary with the way the cable runs at a given time. In addition, during production it is necessary to change the length of bare fibre coming out at the end to gain access during block fitting and polishing. Some spare fibre is also required to be used during the polishing, including some contingency for blocks that have to be cut off and replaced if they get damaged or fail the quality control.

Strain relief boxes at both sides of the 38.1m long main conduit cater to these requirements. In case of the spectrograph side box depicted in Figure 5 and Figure 6, the 53 fibres of each conduit are leaving the 4mm Miniflex conduit, do a loop on a plastic tray and disappear in a new conduit of 5mm diameter to be guided to the slit. The conduit with the three sky fibres ends in the strain relief box, leaving the cleaved ends in a connector plate on its side so the sky signal can be fed into SMA-connectorised fibres that can be plugged in from outside the box. The fibres can now be pushed in or pulled out of the conduit at the slit end. This end is connected with the box by another 4.1m long Adaptaflex cable in which the now ten Miniflex conduits of 3.1mm inner and 5mm outer diameter run in a parallel, not stranded manner without a tensile element. Black Miniflex conduits have been chosen for this stretch to suppress stray light entering the sensitive slit area of the spectrograph.

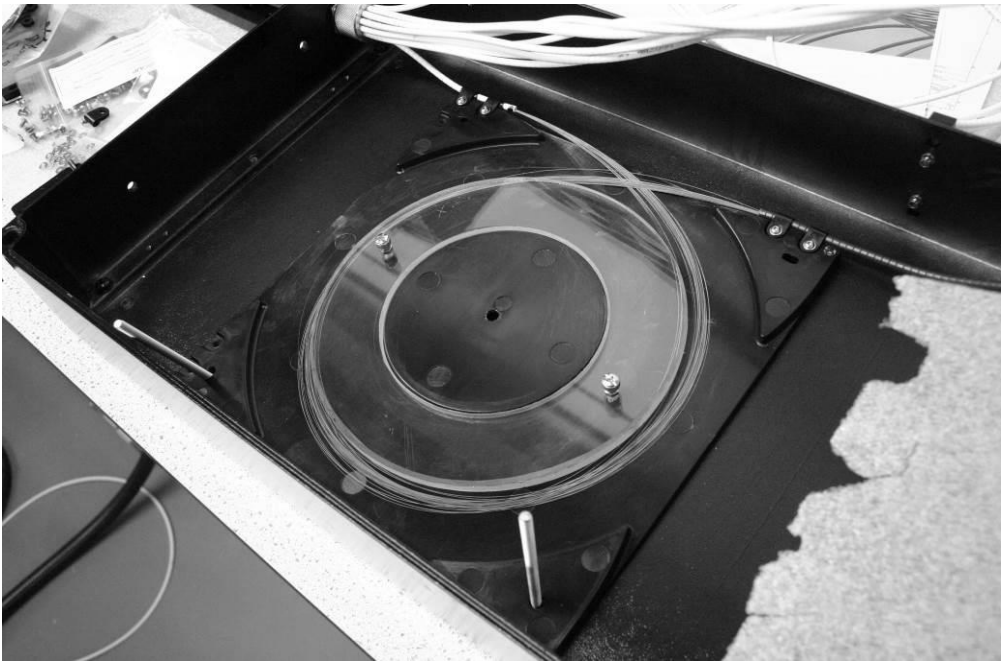

**Figure 5: Fibre loop in the strain relief box. The Perspex template is removed after the loop has been laid.**

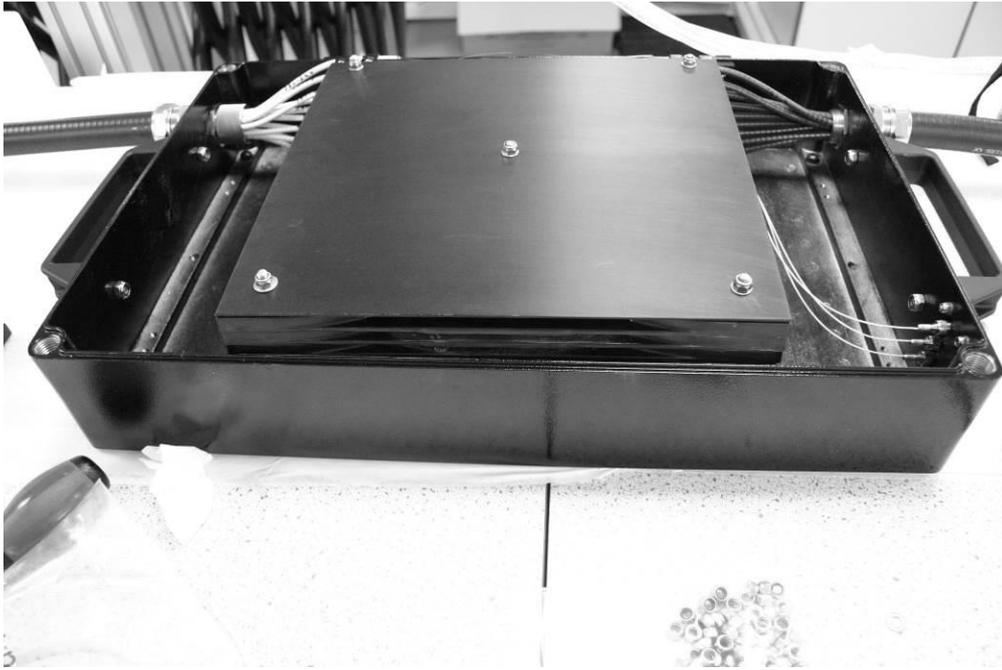

**Figure 6: Fully populated strain relief box before closoing. The top tray carries the three sky fibres which go through protective furcation tubing before terminating at the three SMA feedthroughs at the lower right.**

## 4. ATTACHING THE FIBRE SLIT

The fibres now protrude from the Miniflex tubing. To cater for their different length dictated by the design of the mechanical slit hardware (see Figure 12), a template is used to cut the fibres for each block to an individual length and to protect the fibres with PTFE tubing that helps to guide them through the slit hardware. The fibres of each block (25+spares) share a common PTFE tube, so that two PTFE tubes fit into the end of one M2FX conduit. At this stage also a life test is undertaken to verify that all optical fibres are intact throughout the cable feed.

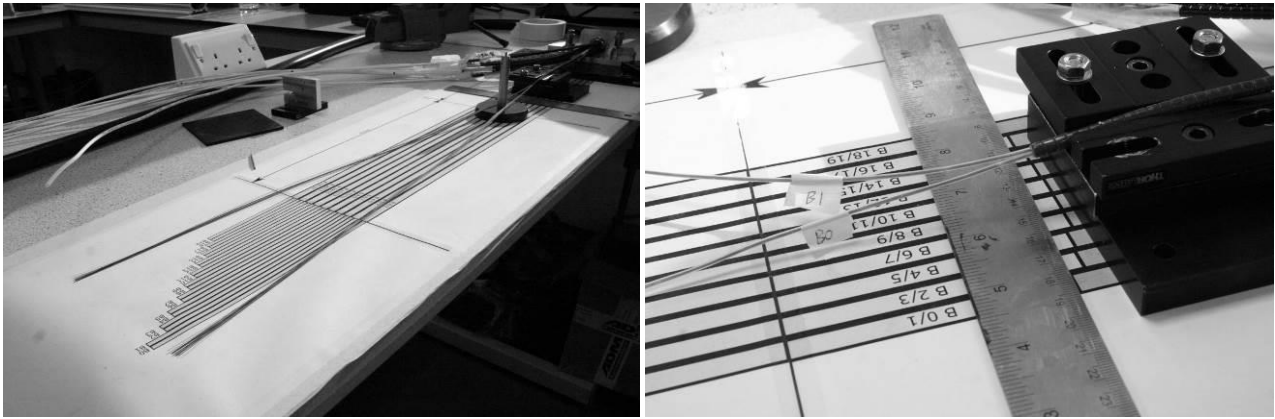

**Figure 7: Terminating the slit side. Left: Cutting the fibres to length using a template. Right: Terminated and labelled fibres in their protective PTFE tubing.**

The fibre ends are attached to 6mm wide v-groove blocks which are then polished and fitted with a broadband AR coated window. The details about the slit block manufacture are provided in Tyas et al, [4]. Once the cable has the blocks fitted, each block is attached to the common slit backing plate and the positions and pointing directions are verified before each block is fixed in place by UV curing adhesive.

**4.1 Attaching the blocks to the slit backing plate**

The 20 slit blocks are bearing 25 parallel fibres on a pitch of 230μm each. As the slit is curved by a nominal radius of 468.3mm, the blocks have to follow this curvature while minimizing the impact on focus position and pointing direction. The specifications for the block positions and orientation are listed in Table 1.

**Table 1: Position requirements for the slit blocks**

| Requirement | Value | Comments |
|---|---|---|
| Grouping of fibres | 20 blocks of 25 fibres | Segmented slit approximating curvature, using flat blocks of parallel fibres |
| Slit radius of curvature | 486.3 +/- 3mm convex | |
| Aperture separation | 230 ± 10 μm in block, 556 ± 50 μm between blocks | |
| Pointing accuracy | 0.1° (dispersion direction) 0.35° (cross dispersion) | Cross dispersion tolerance considers slit curvature approximation by parallel fibres |
| Lateral tolerance along slit inside each block between adjacent blocks | ± 10 μm ± 25 μm | Affects extraction of adjacent spectra |
| Lateral tolerance in dispersion direction | ± 50 μm | Will be compensated by wavelength calibration |
| Diversion from curvature along slit | ± 20 μm | Includes systematic deviation caused by flat blocks following curvature |

The assembly method chosen breaks down to the following steps:
- Laying the fibre bundles out in sequence to verify by a template that each necessary length can be reached. It is important that there are no crossovers as this cannot be corrected for once the blocks are on the backing plate.
- Installing the slit backing plate in a jig that contains datum pins.
- Attaching a datum plate where each slit block pushes against two wire-eroded notches to assure the position and pointing direction is correct (Figure 8 top left).
- A 70μm thick wire is used at a defined position on a glued down block to provide the correct distance for the next one. This wire is lowered from a separate arm from the top.
- Cleaning the block and the plate with high purity isopropylic alcohol.

- Application of a small and reproducible amount of UV curing glue. We dispense low outgassing Norland NOA88 optical glue using a Nordson EFD Performus IV dispenser to attach a reproducible amount of glue to the plate.
- Putting the block into position. A foam-padded weight on the PTFE conduit suppresses any twist. While the glue provides some adhesion to keep the block down, it can be moved into position. A post with a soft nylon tip is then lowered onto the block to keep it down (Figure 8 top right).
- Verification of position using two small microscopic cameras including measurement of the distance between the edge fibres of the new block and the block installed previously. The cameras image the scene from the front and the top, so the datum contacts can be checked as well. The distance of the edge fibres of both blocks is measured using a metrology feature of the camera software to assure it is within $556\pm50\mu m$.
- A further verification of the pointing direction is performed by back-illumination of the central block fibre by a laser. The annulus of light exiting the fibre block is observed on a target screen where the expected positions are plotted as circles.
- After all observations indicate a compliant block position and orientation, UV light of an LED UV curing source is applied for three minutes to fix the block in position.

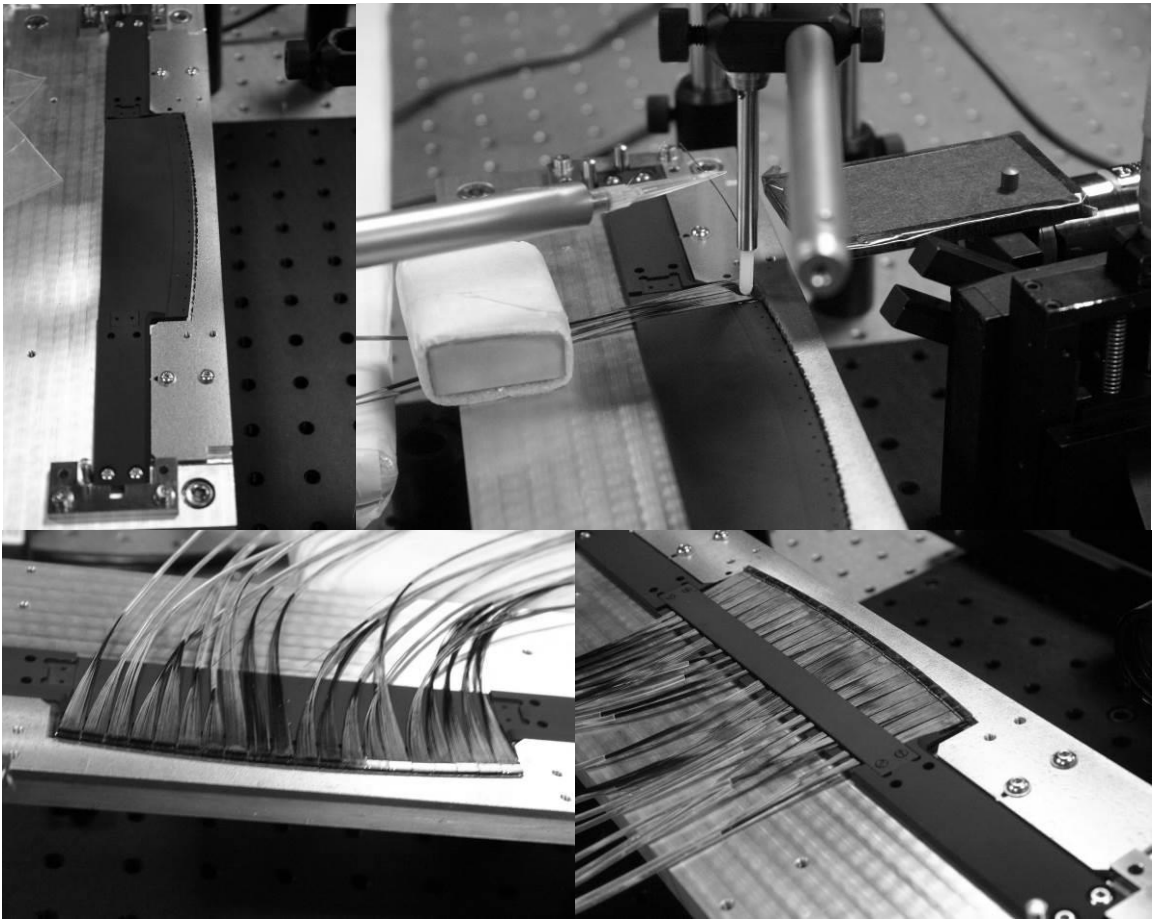

**Figure 8: Slit backing plate population. Top left: Slit in fixture with datum jig in front. Top right: Attachment of first block after glue deposition. Bottom left: All blocks installed, bottom right: Protective plate installed to provide strain relief when handling the backing plate.**

These steps have to be repeated until the full slit is assembled. Once finished, a thin back plate is used to cover the fibre bundles where they leave the blocks. This prevents any accidental bending of the fibres with respect to their fixed blocks which could result in fibre breakage.

### 4.2 Positional verification

At this stage a fully automated 3-axis microscope is used (Figure 9). While diffusely back-illuminating the fibres, the machine automatically focuses onto each fibre and records the position in three spatial dimensions. A separate calibration jig is used to transform the coordinates into an absolute system that links the situation at the assembly jig to the reference points on the final slit mechanical hardware. Any offset can be adjusted on the slit hardware separately and measured with a coordinate measurement machine, without the presence of the delicate fibres.

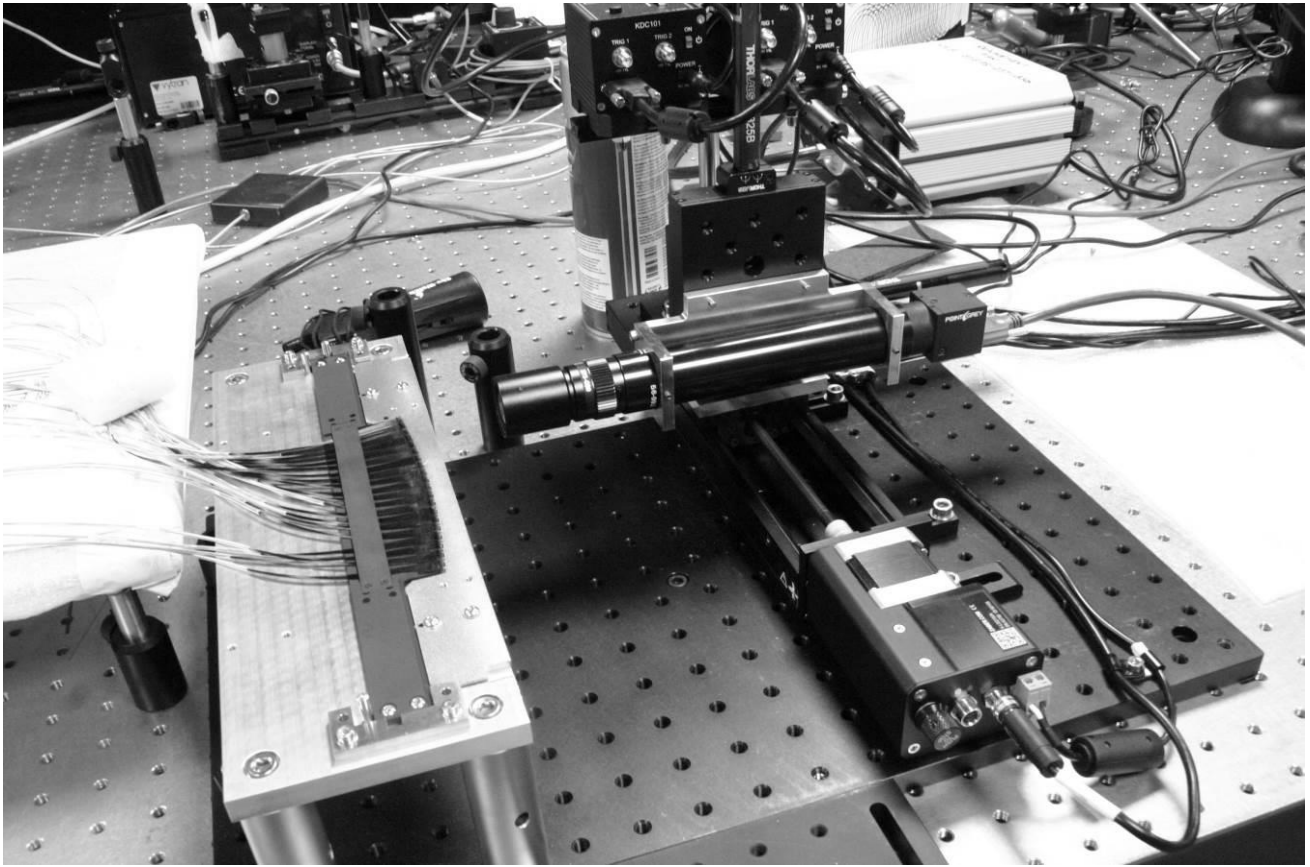

**Figure 9: The 3d stage with microscopic camera used for the automated metrology of all 500 apertures of the slit.**

### 4.3 Completion of slit assembly

The slit backing plate is mounted inside the parallel beam of the collimator that feeds the spectrograph, and a box-like framework is required to minimise vignetting. The whole slit mechanism is able to be aligned in all three spatial directions and tip/tilt by using fine threaded adjusters. The slit unit can be inserted into the spectrograph reproducibly while assuring the slit is not damaged as it slides into a slot ground into the dichroic filter that causes the first wavelength channel separation in the spectrograph. While the slit backing plate receives the blocks, the slit hardware is assembled as seen in Figure 10 and aligned on a coordinate measurement machine (Mitutoyo Chrysta Apex C) to assure the position of the three datum cylinders that the slit backing plate references to is known in space. Any offset caused by the residual curvature of the slit backing plate can be incorporated at this step, as the slit backing plates have been measured for flatness a step before and where necessarey carefully bent back to be straight within 100µm. This process is verified with a Taylor Hobson Talysurf profilometer.

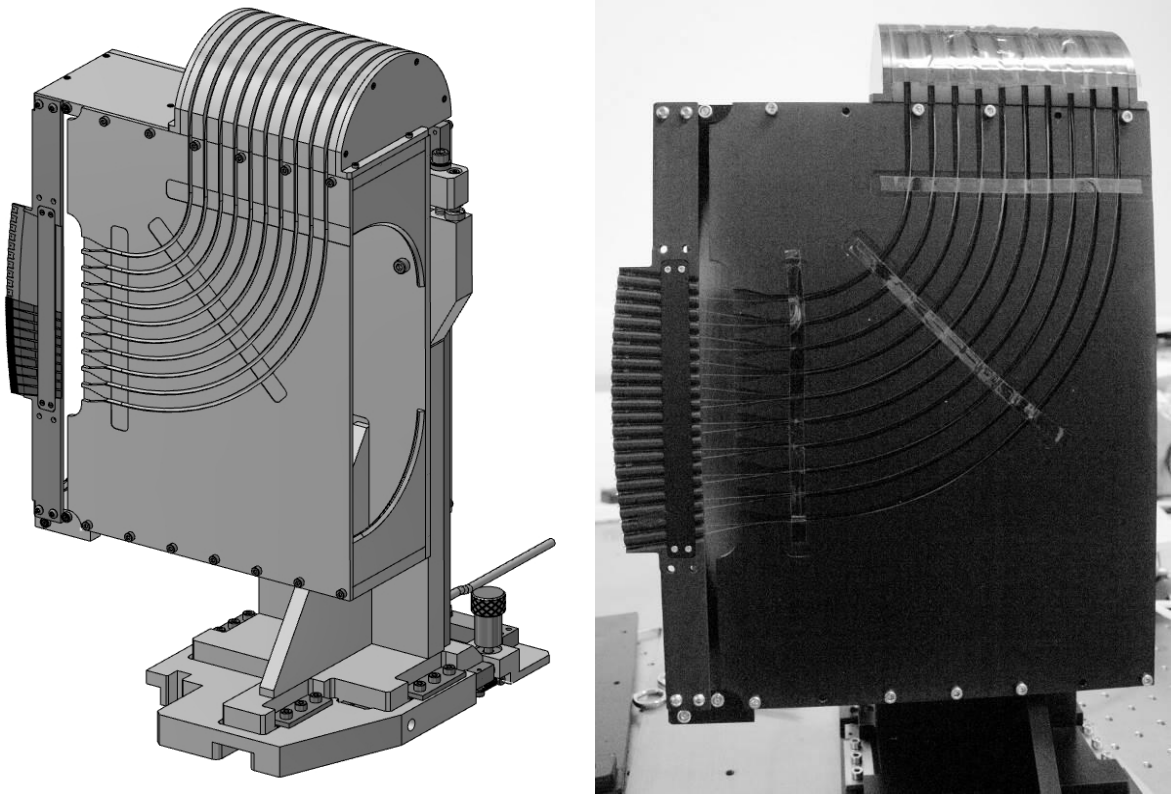

**Figure 10: The slit with its "race track" design minimizes vignetting losses in the collimated beam.**

Extreme care has to be taken when mating the populated slit backing plate with the slit hardware. Some steps of this process can be seen in Figure 11. After the slit backing plate is mounted to three datum cylinders that act as reference to the coordinate metrology mentioned above, each pair of blocks share the same "race track" while being guided out of the optical path and around to the point where the conduits break out of the main conduit. The tracks are machined T-slots which enable the lateral feed of one conduit at a time, with the second PTFE conduit effectively locking the first one into place so that both conduits do not jump out of the slot again. After another semi-circular bend on the top of the assembly, the conduits leave the T-slot to be held into place by various retaining brackets which either hold the PTFE or the Miniflex conduit which the PTFE tubing terminates in. Finally the cables re-arrange from a planar to a circular layout before they disappear in the main cable. While assembling this, the length of the fibres and the PTFE tubing coming out of the black Miniflex material can be varied, as the strain relief box provides the give or take for each fibre bundle. Once the slit is completely assembled, the focal ratio degradation is measured again as it provides a final reading of the whole cable downstream from the point where it will be later fusion-spliced to the PFA assembly.

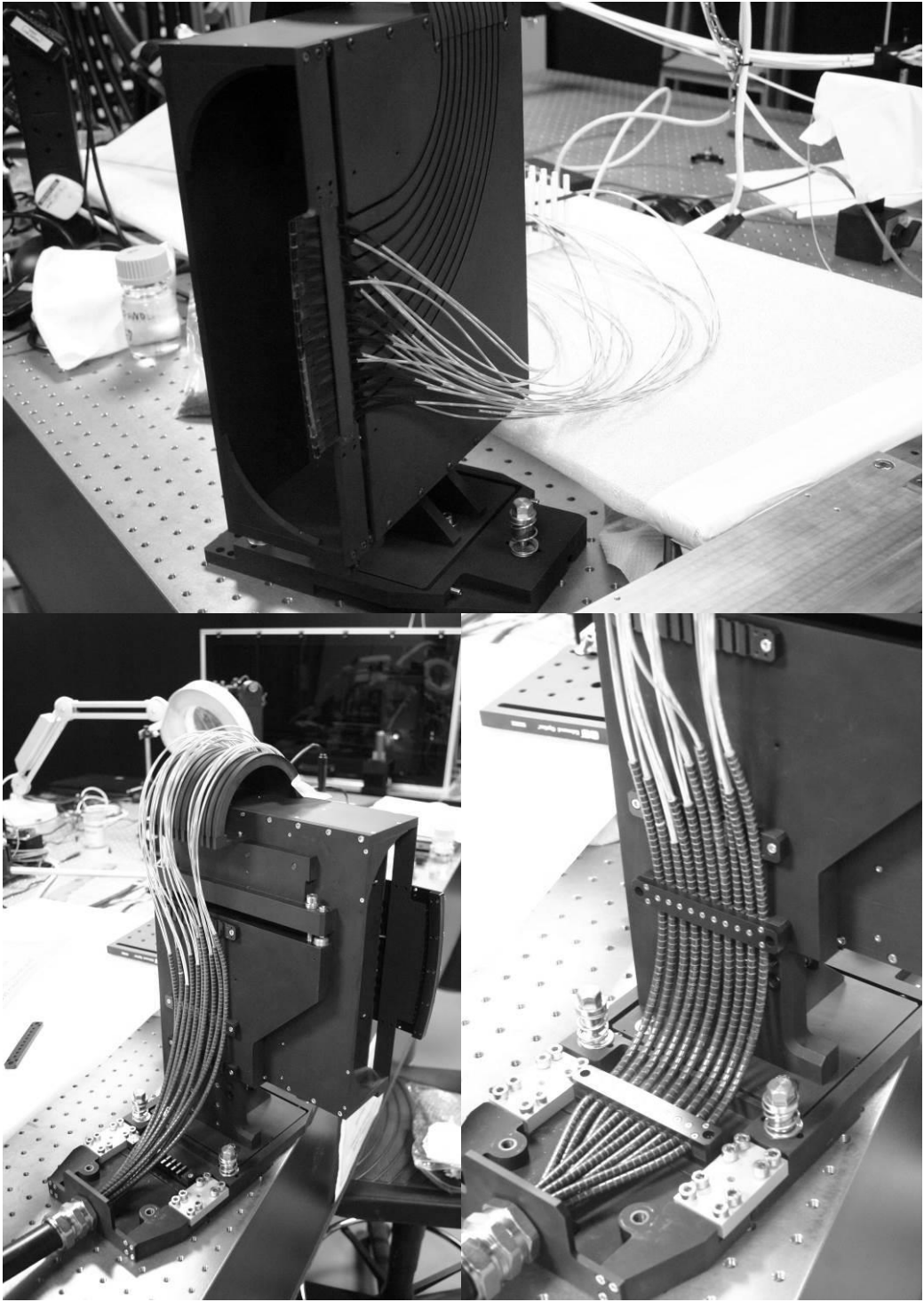

**Figure 11: Integration of the populated slit backing plate into the slit hardware. Top: Backing plate installed, preparing to lay fibres in PTFE into the grooves of the "race track"; bottom left: Fibres during race track integration; bottom right: Fixing the conduits in place using retaining brackets.**

## 4.4 FRD verification

A lever mechanism is used as shown in Figure 12 to illuminate the blocks with a parallel laser beam at an angle near 7.3° corresponding to the opening angle of the f/3.9 beam from the telescope. If the input focal ratio deviates from that figure, a compensation term is applied to the resulting ring wdth. The pivot point of the lever is identical to the radius of curvature of the slit (468.3mm), and the tip of the lever carries a fibre-fed SMA collimator that provides a collimated beam of 525nm light that feeds one slit block at a time. The lever can be re-positioned and clamped in place, while the mechanism is set up so that in case of a failure no parts can touch the slit itself which is located about 20mm in front of the collimator lens. Illuminating a whole block with parallel light allows picking any sub-sample of fibres that get cleaved at the PFA end of the cable, and subsequently FRD measured. Figure 13 shows the result of 100 fibres measured at the first delivered science cable. The mean FWHM of the annulus recorded when illuminating the block with f/3.9 is around one degree, which indicates compliant focal ratio degradation.

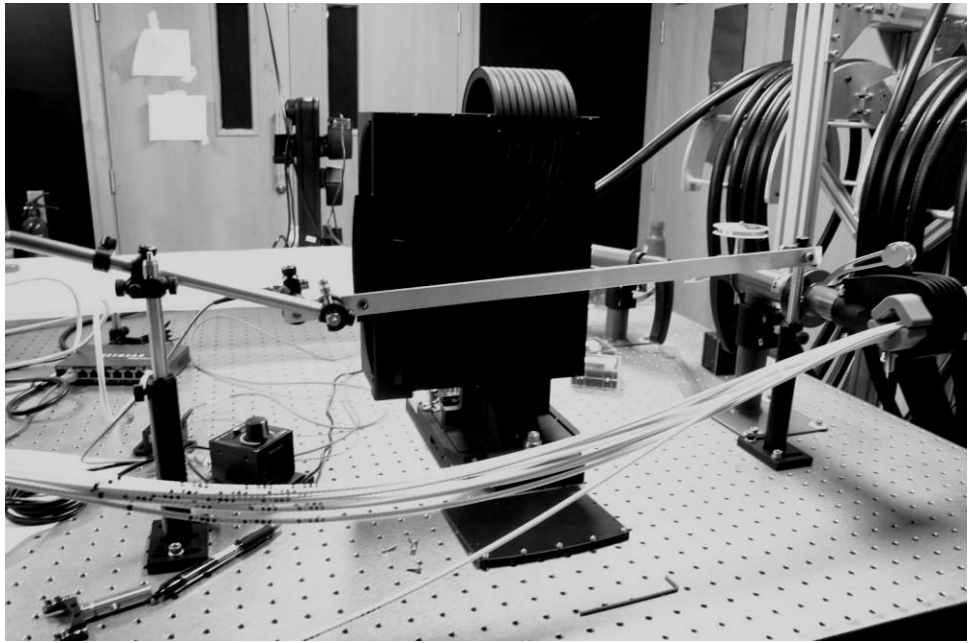

**Figure 12: Focal ratio degradation metrology at the finished slit: The lever mechanism uses a fibre-fed SMA collimator to inject parallel light at the required angle into one block at a time. The lever assures that the collimator follows the curvature of the slit.**

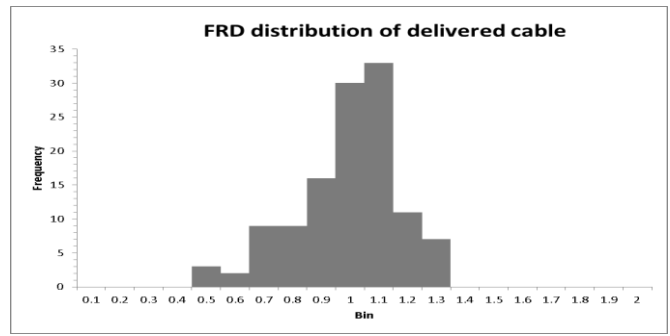

**Figure 13: Histogram of Focal Ratio Degradation performance of a sub-sample of 100 fibres (5 per block) of Science Cable 0, as measured with the lever setup after full assembly of the slit.**

## 5. OUTLOOK

Currently the production of the science cables is in full swing, with completion of ten plus two spare cables expected in the autumn of 2018. The finished cables are shipped to the LNBL in Berkeley, where they are being connected with their PFA ends.

## ACKNOWLEDGEMENTS

This research is supported by the Director, Office of Science, Office of High Energy Physics of the U.S. Department of Energy under Contract No. DE–AC02–05CH1123, and by the National Energy Research Scientific Computing Center, a DOE Office of Science User Facility under the same contract; additional support for DESI is provided by the U.S. National Science Foundation, Division of Astronomical Sciences under Contract No. AST-0950945 to the National Optical Astronomy Observatory; the Science and Technologies Facilities Council of the United Kingdom; the Gordon and Betty Moore Foundation; the Heising-Simons Foundation; the National Council of Science and Technology of Mexico, and by the DESI Member Institutions. The authors are honored to be permitted to conduct astronomical research on Iolkam Du'ag (Kitt Peak), a mountain with particular significance to the Tohono O'odham Nation. This work has been funded by the STFC grant number ST/M002675/1.

A note of thanks goes to the PFS collaboration and to PPC Broadband Fiber Ltd. for enabling the cable stranding by using the strander developed by PPC for the PFS project.